\def\nk{n_{\rm b}}

\def\snf{\sin f}
\def\csf{\cos f}

\def\Pb{P_{\rm b}}

\def\rfr#1{Equation\,(\ref{#1})}
\def\rfrs#1#2{Equations\,(\ref{#1})--(\ref{#2})}
\def\Rfr#1{Equation\,(\ref{#1})}

\def\derp#1#2{\rp{\partial{#1}}{\partial{#2}}}
\def\dert#1#2{\frac{{{\textrm{d}}}{#1}}{{{\textrm{d}}}{#2}}}

\def\eqi{\begin{equation}}
\def\eqf{\end{equation}}
\def\eqia{\begin{eqnarray}}
\def\eqfa{\end{eqnarray}}

\def\rp#1#2{{#1\over#2}}
\def\lb#1{\label{#1}}

\def\bds#1{\boldsymbol{#1}}

\def\ton#1{\left(#1\right)}
\def\qua#1{\left[#1\right]}
\def\grf#1{\left\{#1\right\}}
\def\ang#1{\left\langle #1\right\rangle}
\documentclass[onecolumn]{aastex}

\usepackage{morefloats}
\usepackage[title]{appendix}
\usepackage{hyperref,textcomp}
\usepackage{booktabs}
\usepackage[table,xcdraw]{xcolor}
\usepackage{multirow}
\usepackage{rotating,tabularx}
\usepackage{float}
\usepackage{enumerate}
\usepackage{rotating}
\usepackage[polutonikogreek,english]{babel}
\usepackage{amsmath,starfont,textgreek,w-greek,wasysym}
\usepackage[flushleft]{threeparttable}
\usepackage{amsthm}
\usepackage{amscd,lineno}
\usepackage{amssymb,dsfont}
\usepackage{graphicx,epsfig}
\usepackage{txfonts}
\usepackage{xr-hyper}

\RequirePackage{color}

\newcommand{\emaila}{lorenzo.iorio@libero.it}

\allowdisplaybreaks[1]

\begin{document}

\title{Post-Newtonian effects on some characteristic timescales of transiting exoplanets}

\shortauthors{L. Iorio}

\author{Lorenzo Iorio\altaffilmark{1} }
\affil{Ministero dell'Istruzione, dell'Universit\`{a} e della Ricerca
(M.I.U.R.)
\\ Viale Unit\`{a} di Italia 68, I-70125, Bari (BA),
Italy}

\email{\emaila}

\begin{abstract}
Some measurable characteristic timescales $\grf{t_\mathrm{trn}}$ of transiting exoplanets are investigated in order to check preliminarily if their cumulative shifts over the years induced by the post-Newtonian (pN) gravitoelectric (Schwarzschild) and gravitomagnetic (Lense-Thirring) components of the stellar gravitational field are, at least in principle, measurable. Both the primary (planet in front of the star) and the secondary (planet behind the star) transits are considered along with their associated characteristic time intervals: the total transit duration $t_D$, the ingress/egress transit duration $\uptau$, the full width at half maximum primary transit duration $t_H$, and also the time of conjunction $t_\mathrm{cj}$. For each of them, the net changes per orbit $\ang{\Delta t_D},\,\ang{\Delta\uptau},\,\ang{\Delta t_H},\,\ang{\Delta t_\mathrm{cj}}$ induced by the aforementioned pN accelerations are analytically obtained; also the Newtonian effect of the star's quadrupole mass moment $J_2^\star$ is worked out. They are calculated for a fictitious Sun-Jupiter system in an edge-on elliptical orbit, and the results are compared with the present-day experimental accuracies for the HD 286123 b  exoplanet. Its pN gravitoelectric shift $\ang{\Delta t_\mathrm{cj}^\mathrm{1pN}}$ may become measurable, at least in principle, at a $\simeq 8\times 10^{-5}$ level of (formal) relative accuracy after about 30 years of continuous monitoring corresponding to about 1000 transits. Systematics like, e.g.,
confusing time standards, neglecting star spots, neglecting clouds, would likely deteriorate the actual accuracy.  The method presented is general enough to be applied also to modified models of gravity.
\end{abstract}


\section{Inroduction}
The term exoplanets refers to major natural bodies orbiting stars other than our Sun \citep{2010exop.book.....S,2018haex.bookE....D,2018exha.book.....P}.
Among them, transiting exoplanets \citep{2010trex.book.....H} are detected by monitoring the drop in the collected electromagnetic flux  when they pass in front of their parent star in what is called as primary transit or primary eclipse \citep{Winn2011}. Since it is the combined flux from both the star and the planet which is actually measured, a further reduction of it occurs also when the planet passes behind the star in what is called as secondary transit or secondary eclipse \citep{Winn2011}. In both cases, some characteristic timescales $\grf{t_\mathrm{trn}}$ are measured: the total transit duration $t_D$, the ingress/egress transit duration $\uptau$, the full width at half maximum primary transit duration $t_H$, and also the times of inferior and superior conjunctions $t_\mathrm{cj}$ \citep{2019arXiv190709480E}.
The collected data records, spanning months or even years, usually cover a large number of transits $N_\mathrm{tr}$  since most of the detected transiting exoplanets are close to their parent stars. Suffice it to say that K2--137 b \citep{2018MNRAS.474.5523S}, discovered in 2017, is only $0.0058\,\mathrm{au}$ from its star and has an orbital period $\Pb$ as short as $4.3\,\mathrm{h}$; in principle, more than $N_\mathrm{tr} = 10000$ transits are available nowadays for it. Furthermore, at the time of writing, the number of confirmed transiting exoplanets returned by
the website http://exoplanet.eu/ amounts to more than 3500. It is destined to further increase thanks to the (re)analysis of large data sets collected by numerous past, present and future dedicated Earth-based and spaceborne surveys like CoRot (Convection, Rotation and planetary Transits) \citep{2006ESASP1306...33B}, TESS (Transiting Exoplanets Survey Satellite) \citep{tess14}, Kepler \citep{2008IAUS..249...17B}, K2 \citep{2014PASP..126..398H}, KELT (Kilodegree Extremely Little Telescope) \citep{2007PASP..119..923P}, UKIRT (United Kingdom Infrared Telescope) \citep{UKIRT}, NGTS (Next-Generation Transit Survey) \citep{2013EPJWC..4713002W}, PLATO (PLAnetary Transits and Oscillations of stars) \citep{2016AN....337..961R}.

Purpose of this work is to devise a general analytical method to calculate the net variations per orbit $\ang{\Delta t_\mathrm{trn}}$ of the aforementioned timescales $\grf{t_\mathrm{trn}}$ induced by any post-Keplerian (pK) dynamical features, of Newtonian, post-Newtonian (pN), or even exotic origin, and preliminarily investigate their potential detectability over the years thanks to their cumulative nature by looking at the current level of accuracy $\sigma_{t_\mathrm{trn}}$ in measuring $\grf{t_\mathrm{trn}}$.
In this paper, the focus is on the known pN Schwarzschild and Lense-Thirring (LT) features \citep{SoffelHan19}, and on the pK Newtonian effect of the quadrupole mass moment $J_2^\star$ of the host star. Indeed, if one is interested in pN gravity, $J_2^\star$ would act as a source of systematic error whose biasing effect has to be modelled as accurately as possible.
For previous studies on general relativity in exoplanets and the possibility of using them as probes to measure its effects on them, see \citet{2006ApJ...649..992A,2006ApJ...649.1004A,2006IJMPD..15.2133A,2006NewA...11..490I,2008ApJ...685..543J,2008MNRAS.389..191P,2011MNRAS.411..167I,
2011Ap&SS.331..485I,2012Ap&SS.341..323L,2013RAA....13.1231Z,2016MNRAS.455..207I,2016MNRAS.460.2445I,2019A&A...628A..80B,2021MNRAS.505.1567A,2021E&ES..658a2051G}
Modified models of gravity can be treated with the method set out here as well. By confronting the resulting expressions for the induced net shifts $\ang{\Delta t_\mathrm{trn}}$ of the relevant timescales $\grf{t_\mathrm{trn}}$ with the current formal accuracy in measuring them can yield preliminary insights into the possibility of constraining the models' key parameters. It is outside the scopes of the present work; for some previous studies, see, e.g., \citet{2010OAJ.....3..167I,2014MNRAS.438.1832X,2017PhLB..769..485V,2020JCAP...06..042R,2021PhRvD.104h4097K}.

The paper is organized as follows. In Section\,\ref{metodo}, the general calculational scheme is outlined. Section\,\ref{primary} deals with the primary transit and related timescales. In particular, Section\,\ref{trdr} hosts a calculation of the (Keplerian) total transit duration $t_D$ which is subsequently used to work out its pN and classical pK average shifts (Sections\,\ref{TGE}\,to\,\ref{TJ2}). Some numerical evaluations for a fictitious Sun--Jupiter system with an edge-on eccentric orbit and a confrontation with the present-day experimental accuracy for the exoplanet HD 286123 b are made in Section\,\ref{misura}.
The ingress/egress transit duration $\uptau$ is treated in Section\,\ref{ingr}, while Section\,\ref{fwhm} is devoted to the full width at half maximum primary transit duration $t_H$. The time of inferior conjunction $t_\mathrm{cj}$ is the subject of  Section\,\ref{TC}. In Section\,\ref{secondary}, the same calculation are repeated for the case of the secondary eclipse. In Section\,\ref{concludi},  the main results are presented, and the conclusions are offered.
\section{Outline of the proposed method}\label{metodo}
Let $\mathfrak{O}$ be some observable quantity for a gravitationally bound binary system made of two bodies $\textrm{A},~\textrm{B}$  experiencing, among other things, an extra--acceleration $\bds A$, of arbitrary physical origin, which can be considered as small with respect to the usual Newtonian inverse-square law monopole. In the case of exoplanetary systems, $\mathfrak{O}$ could be, e.g.,  the transit duration $t_D$, the ingress/egress transit duration $\uptau$, the full width at half maximum primary transit duration $t_H$, the time(s) of conjunction(s) $t_\mathrm{cj}$, and the radial velocity $V$. Let us assume that $\mathfrak{O}$ can be modelled in terms of some explicit function $F\ton{a,~e,~I,~\Omega,~\omega,~f}$ of the Keplerian orbital elements $a,~e,~I,~\Omega,~\omega,~f$ which are the semimajor axis, the eccentricity, the inclination, the longitude of the ascending node, the argument of pericentre, and the true anomaly, respectively; its instantaneous pK change $\Delta F\ton{f}$ induced by $\bds A$ can be straightforwardly obtained as
\begin{equation}
\Delta F\ton{f} = \sum_\zeta\derp{F}{\zeta}\,\Delta\zeta\ton{f},\,\zeta=a,~e,~I,~\Omega,~\omega,~f\label{DF},
\end{equation}
In \rfr{DF}, the instantaneous variations of all the Keplerian orbital elements, apart from the true anomaly $f$, are  worked out, to the first order in $A$, as
\begin{equation}
\Delta\zeta\ton{f} = \int_{f_0}^f\dert{\zeta}{t}\,\dert{t}{f^{'}}\,\mathrm{d}f^{'},~\zeta=a,~e,~I,~\Omega,~\omega,\label{Dk}
\end{equation}
where $f_0$ is the true anomaly at some initial epoch $t_0$.
In \rfr{Dk},
the time derivatives $d\zeta/dt$ are to be computed by calculating, for a given disturbing acceleration $\bds A$, the right-hand-sides of the usual Gauss equations of the variation of the Keplerian orbital elements \citep{Nobilibook87,1991ercm.book.....B,2003ASSL..293.....B,2008orbi.book.....X,2011rcms.book.....K}
\begin{align}
\dert a t \lb{dadt}& = \rp{2}{\nk\,\sqrt{1-e^2}}\,\qua{e\,A_\mathrm{R}\,\snf + \ton{\rp{p}{r}}\,A_\mathrm{T}}, \\ \nonumber\\
\dert e t \lb{dedt} & = \rp{\sqrt{1-e^2}}{\nk\,a}\,\grf{A_\mathrm{R}\,\snf + A_\mathrm{T}\,\qua{\csf + \rp{1}{e}\,\ton{1-\rp{r}{a}} }}, \\ \nonumber\\
\dert I t & = \rp{1}{\nk\,a\,\sqrt{1-e^2}}\,A_\mathrm{N}\,\ton{\rp{r}{a}}\,\cos u, \\ \nonumber\\
\dert \Omega t & = \rp{1}{\nk\,a\,\sin I\,\sqrt{1-e^2}}\,A_\mathrm{N}\,\ton{\rp{r}{a}}\,\sin u, \\ \nonumber\\
\dert \omega t \lb{dodt} & = \rp{\sqrt{1-e^2}}{\nk\,a\,e}\,\qua{-A_\mathrm{R}\,\csf + A_\mathrm{T}\,\ton{1 + \rp{r}{p}}\,\snf} - \cos I\,\dert\Omega t
\end{align}
onto the unperturbed Keplerian ellipse
\begin{equation}
r=\rp{p}{1+e\cos f},\label{rKep}
\end{equation}
and
\begin{equation}
\dert{t}{f} = \rp{\ton{1-e^2}^{3/2}}{\nk\,\ton{1+e\,\cos f}^2}.\lb{dfdt}
\end{equation}
In \rfrs{dadt}{dfdt}, $A_\mathrm{R},\,A_\mathrm{T},\,A_\mathrm{N}$ are the radial, transverse and normal components of $\bds A$, respectively, obtained by projecting it onto the unit vectors $\mathbf{\hat{\mathrm{R}}},\,\mathbf{\hat{\mathrm{T}}},\,\mathbf{\hat{\mathrm{N}}}$ of the aforementioned directions,  $r$ is given by \rfr{rKep}, $\mu$ is the gravitational parameter of the system at hand given by the product of the Newtonian constant of gravitation $G$ times the sum $M_\mathrm{tot} = M_\mathrm{A} + M_\mathrm{B}$ of the masses of the bodies A and B, $\nk\doteq\sqrt{\mu/a^3}$ is the unperturbed Keplerian mean motion, $u\doteq \omega + f$ is the argument of latitude, and $p\doteq a\,\ton{1-e^2}$ is the semilatus rectum.

The pK change $\Delta f\ton{f}$ of the true anomaly $f$ is subtler, and involves the change $\Delta\mathcal{M}\ton{f}$ of the mean anomaly $\mathcal{M}$, whose calculation requires care, as it will be shown below. According to Equation (A.6) of \citet{1993CeMDA..55..209C}, $\Delta f$ can be written
\begin{equation}
\Delta f\ton{f} = \ton{\rp{a}{r}}\,\qua{\sin f\,\ton{ 1 + \rp{r}{p} }\,\Delta e\ton{f} + \sqrt{1-e^2}\,\ton{\rp{a}{r}}\,\Delta{\mathcal{M}}\ton{f}}.\label{Dfana}
\end{equation}
While the calculation of $\Delta e\ton{f}$ is straightforward as per \rfr{Dk} with \rfr{dedt} and \rfr{dfdt}, $\Delta\mathcal{M}\ton{f}$ is more complicated to be worked out; see, e.g., \citet{1991ercm.book.....B,1989racm.book.....S}.
As shown in \citet{2017EPJC...77..439I}, the change of the mean anomaly can be obtained, e.g., as
\begin{equation}
\Delta\mathcal{M}\ton{f}=\Delta\eta\ton{f} + \int_{t_0}^t\Delta\nk\ton{t^{'}}\,\mathrm{d}t^{'},\label{DM}
\end{equation}
where the variations of  the mean anomaly at epoch $\eta$ and of the mean motion $\nk$ are
\begin{align}
\Delta\eta\ton{f} \label{Deta}&= \int_{f_0}^f\dert{\eta}{t}\,\dert{t}{f^{'}}\,\mathrm{d}f^{'},\\ \nonumber \\
\int_{t_0}^t\Delta\nk\ton{t^{'}}\,\mathrm{d}t^{'} \label{intDn}& = -\rp{3}{2}\rp{\nk}{a}\int_{f_0}^f\Delta a\ton{f_0,~f^{'}}\,\dert{t}{f^{'}}\,\mathrm{d}f^{'},
\end{align}
respectively.
In \rfr{Deta}, it is  \citep{Nobilibook87,1991ercm.book.....B,2003ASSL..293.....B}
\begin{equation}
\dert{\eta}t = - \rp{2}{\nk a}\,A_{\rho}\,\ton{\rp{r}{a}} -\rp{\ton{1-e^2}}{\nk a e}\,\qua{ -A_{\rho}\,\csf + A_{\tau}\,\ton{1 + \rp{r}{p}}\,\snf },\label{detadt}
\end{equation}
while $\Delta a\ton{f}$ entering \rfr{intDn} is worked out according to \rfr{Dk} with \rfr{dadt} and \rfr{dfdt}.
Depending on the specific perturbing acceleration $\bds A$ at hand, the calculation, especially of  \rfr{intDn}, can be, in general, rather unwieldy.
In this case, it may turn out to be convenient using the eccentric anomaly $E$ instead of $f$ as fast variable of integration by means of
\begin{align}
r \lb{rE}& = a\,\ton{1-\cos E}, \\ \nonumber \\
\sin f \lb{sfE} & = \rp{\sqrt{1-e^2}\,\sin E}{1 - e\,\cos E}, \\ \nonumber \\
\cos f \lb{cfE} & = \rp{\cos E - e}{1 - e\,\cos E}, \\ \nonumber \\
\dert{t}{E} \lb{dtdE} & = \rp{1 - e\,\cos E}{\nk}
\end{align}
in \rfr{Dk}, \rfrs{dadt}{dodt}, and \rfrs{Deta}{intDn}.


The net change per orbital revolution  $\ang{\Delta F}$ is obtained from \rfr{DF} with the substitution $f\rightarrow f_0 + 2\uppi$, or $E\rightarrow E_0 + 2\uppi$, in it; it turns out that, in general, $\ang{\Delta F}$ may depend on $f_0$ or on $E_0$.

The calculational scheme outlined here is general enough to be applied to whatsoever extra-acceleration $\bds A$, provided it is small enough to be treated perturbatively.  As such, although in the next Sections it will be used with standard Newtonian and pN dynamical effects, in principle, it can be extended also to any alternative models of gravity to perform sensitivity analyses and put preliminary constraints on their relevant parameters; it is outside the scopes of the present paper.
\section{The primary transit: when the planet passes in front of the star}\lb{primary}
Let p be a planet of mass $M_\mathrm{p}$ and radius $R_\mathrm{p}$ passing in front of a star s of mass $M_\star$ and radius $R_\star$ along an inclined and eccentric orbit. The starlight is partially blocked by the planet, and it is said that the primary transit, or the primary eclipse, occurs; for a general overview of transits and occultations of exoplanets, see, e.g., \citet{Winn2011}, and references therein.
The transit starts at the first instant of contact $t_\mathrm{I}$, when the planet's disk, moving towards the star, becomes externally tangent to the stellar one. Then, at the second instant of contact $t_\mathrm{II}$, the two disks are internally tangent with the planet's disk superimposed to the star's one. At the third instant of contact $t_\mathrm{III}$, the planetary disk begins to leave the stellar one getting internally tangent to it.  The transit  ends at the fourth instant con contact $t_\mathrm{IV}$ when the planetary disk, moving away from the star's one, becomes externally tangent to it.

Some timescales characterizing such a pattern are, actually, measured in data analyses of transiting exoplanets \citep{2019arXiv190709480E}: they are, e.g., the total transit duration $t_D$, denoted as $T_{14}$ by \citet{2019arXiv190709480E}, the ingress/egress transit duration $\uptau$, and the full width at half maximum primary transit duration $t_H$, dubbed as $T_{FWHM}$ by \citet{2019arXiv190709480E}. Also the time of inferior conjunction $t_\mathrm{cj}$, named $T_C$ in \citet{2019arXiv190709480E}, is determined in data reductions. In Sections\,\ref{trdr}\,to\,\ref{TC}, the net shifts per orbit induced by the aforementioned pK accelerations are analytically worked out for each of them.
\subsection{The total transit duration $t_D$}\lb{trdr}
The transit is viewed in the plane of the sky, assumed as reference $\grf{x,\,y}$ plane of an astrocentric coordinate system whose reference $z$ axis is directed towards the observer along the line of sight. In order to obtain a manageable analytical expression for its total duration $t_D$, it is assumed \citep{2008ApJ...689..499C,2008ApJ...678.1407F} that the distance between the planet and its parent star is large enough so that the orbital period $\Pb$ is much longer than $t_D$. Thus, the planetary disk moves across the stellar one along an approximately rectilinear segment at an essentially constant speed which can be assumed equal to that at midtransit $\mathrm{v}_\mathrm{mid}$; in general, the speed along an elliptical orbit is variable since it is
\eqi
\mathrm{v}=\rp{\nk\,a}{\sqrt{1-e^2}}\,\sqrt{1+2\,e\,\cos f+ e^2}.\lb{speed}
\eqf
Furthermore, it can also be assumed that the star-planet separation $r$, generally variable according to \rfr{rKep}, remains substantially unchanged and equal to its value at midtransit.

The coordinates of the center of the planet in the plane of the sky are, in general,
\begin{align}
x_\mathrm{p} \lb{xp}& = r\,\ton{\cos\Omega\,\cos u - \cos I\,\sin\Omega\,\sin u}, \\ \nonumber \\
y_\mathrm{p} \lb{yp}& =r\,\ton{\sin\Omega\,\cos u + \cos I\,\cos\Omega\,\sin u}.
\end{align}
The assumed rectilinear chord of the stellar disk traversed during the transit is parallel to the line of the nodes, i.e. the intersection of the orbital plane with the plane of the sky. Thus, let the reference $x$ axis be aligned just along it, so that $\Omega=0$.
With such a choice, \rfrs{xp}{yp} reduce to
\begin{align}
x_\mathrm{p} \lb{Xp}& = r\,\cos u, \\ \nonumber \\
y_\mathrm{p} \lb{Yp}& = r\,\cos I\,\sin u.
\end{align}

The total transit duration $t_D$ is defined as
\eqi
t_D\doteq t_\mathrm{IV} - t_\mathrm{I}.\lb{tidi}
\eqf
The transit latitude $\alpha_\mathrm{I}$, corresponding to $t_\mathrm{I}$, can be defined by imposing the condition that the segment
\eqi
\ell\,\sin\alpha_\mathrm{I},
\eqf
where
\eqi
\ell\doteq R_\star + R_\mathrm{p}=R_\star\,\ton{1 + \rho},\lb{rhoo}
\eqf
with
\eqi
\rho\doteq \rp{R_\mathrm{p}}{R_\star},\lb{xi}
\eqf
is equal to the $y$ coordinate of p at $t_\mathrm{I}$,
i.e.
\eqi
\ell\,\sin\alpha_\mathrm{I} = y_\mathrm{p}^\mathrm{I}.
\eqf
Thus, from \rfr{Yp}, one reads
\eqi
\sin\alpha_\mathrm{I} = \rp{r_\mathrm{I}\cos I\sin u_\mathrm{I}}{\ell}.\lb{sinfa}
\eqf

Since it is assumed that, during the transit, the planet moves rectilinearly in front of the star without changing their mutual separation, $y_\mathrm{p}$ does not change during $t_D$, and it can be posed equal to its value at midtransit occurring when
\eqi
x_\mathrm{p}=0,
\eqf
i.e., from \rfr{Xp}, for
\eqi
u_\mathrm{mid} = \rp{\uppi}{2},
\eqf
corresponding to
\eqi
f_\mathrm{mid} = \rp{\uppi}{2} -\omega.\lb{fmid}
\eqf
Thus, from \rfr{rKep}, \rfr{sinfa} and \rfr{rhoo}, one has
\eqi
\sin\alpha_\mathrm{I} = \rp{a\,\ton{1-e^2}\cos I}{R_\star\,\ton{1+\rho}\,\ton{1 + e\,\sin\omega}}.\lb{sinalfa}
\eqf

The total transit duration $t_D$ can be expressed as the ratio
\eqi
t_D =\rp{L}{\mathrm{v}_\mathrm{mid}},\lb{tidi}
\eqf
where $L$ is the length of the approximately rectilinear segment traversed from $t_\mathrm{I}$ to $t_\mathrm{IV}$; for an ideal edge-on orbital geometry, i.e. for $I = \uppi/2$, it is
\eqi
L = 2\,\ell,
\eqf
while for an arbitrary orbital inclination it is  given by
\eqi
L= 2\,\ell\,\cos\alpha_\mathrm{I}.\lb{Lcosa}
\eqf
According to \rfr{speed} and \rfr{fmid},  the planet's speed at midtransit is
\eqi
\mathrm{v}_\mathrm{mid} = \rp{\nk\,a}{\sqrt{1-e^2}}\,\sqrt{1 + 2\,e\sin\omega + e^2}.\lb{vmid}
\eqf
It should be remarked that, in the existing literature, \rfr{vmid} is approximated as
\eqi
\mathrm{v}_\mathrm{mid} \simeq \rp{\nk\,a}{\sqrt{1-e^2}}\,\ton{1 + e\,\sin\omega};
\eqf
see, e.g., \citet{2008ApJ...689..499C}. Such an approximation is not adopted in the present work.
Thus, putting together \rfr{tidi}, \rfr{Lcosa}, \rfr{sinalfa} and \rfr{vmid}, one finally has
\eqi
t_D=\rp{2\,R_\star\,\sqrt{1-e^2}}{\nk\,a\,\sqrt{1 + 2\,e\,\sin\omega + e^2}}\,\sqrt{\ton{1+\rho}^2 - b^2},\lb{grossa}
\eqf
where
\eqi
b \doteq \rp{a\,\ton{1-e^2}\,\cos I}{R_\star\,\ton{1 + e\sin\omega}}.\lb{bi}
\eqf

From \rfr{grossa}, it turns out that, by applying \rfr{DF} to it, its net shift per orbit $\ang{\Delta t_D}$ can be obtained straightforwardly by inserting  the averaged variations of the Keplerian orbital elements in the outcome of \rfr{DF}. In the case of \rfr{grossa}, only $a,\,e,\,I,\,\omega$ are involved; for most of the physically relevant perturbing accelerations $\bds A$, only the pericentre $\omega$ and, sometimes, the inclination $I$ undergo non-vanishing averaged changes. More specifically, in a general two-body scenario where the star's spin axis ${\bds{\hat{S}}}_\star$ is arbitrarily oriented in space, $I$ is impacted only by the pK accelerations due to the quadrupole mass moment $J_2^\star$ of the star, of Newtonian origin, and, to the first post-Newtonian (1pN) order, by the stellar angular momentum ${\bds S}_\star$ through the gravitomagnetic LT effect. Instead, the pericentre $\omega$, in addition to the aforementioned long-term effects, is secularly displaced also by the 1pN gravitoelectric acceleration depending only on the masses of the star and the planet \citep{SoffelHan19}. Furthermore, contrary to the inclination, the argument of pericentre is, in principle, affected also by several viable modified models of gravity.
The explicit expressions  of $\ang{\Delta I}$ and $\ang{\Delta\omega}$ calculated for the 1pN gravitoelectric and gravitomagnetic  accelerations and for the effect of $J_2^\star$ can be found in
\citet{2017EPJC...77..439I}, where no a priori assumptions on the spatial orientation of $\bds{\hat{S}}_\star$ are assumed.
The relevant partial derivatives of \rfr{grossa} entering \rfr{DF} are
\begin{align}
\derp{t_D}{I} \lb{dtDdI} & = \rp{a\,\ton{1 - e^2}^{5/2}\,\sin 2I}{\nk\,R_\star\,\ton{1 + e\,\sin\omega}^2\,\sqrt{1 + e^2 + 2\,e\,\sin\omega}\,\sqrt{\ton{1 + \rho}^2 - b^2}}, \\ \nonumber \\
\derp{t_D}{\omega} \lb{dtDdI} & = \rp{2\,e\,\sqrt{1 - e^2}\,\cos\omega\,\qua{-R_\star^2\,\ton{1+\rho}^2\,\ton{1 + e\,\sin\omega}^3 + a^2\,\ton{1 - e^2}^2\,\cos^2 I\,\ton{2 + e^2 + 3\,e\,\sin\omega}}}{\nk\,a\,R_\star\,\ton{1 + e\,\sin\omega}^3\,\ton{1 + e^2 + 2\,e\,\sin\omega}^{3/2}\,\sqrt{\ton{1 + \rho}^2 - b^2}}.
\end{align}
\subsubsection{The 1pN gravitoelectric shift of $t_D$}\lb{TGE}
The 1pN gravitoelectric net change per orbit $\ang{\Delta t^\mathrm{1pN}_D}$ of the total transit duration $t_D$ for two spherically symmetric, static bodies can be calculated according to the strategy outlined in Section\,\ref{metodo} by using \rfr{grossa} in \rfr{DF}.

The resulting outcome is too cumbersome to be explicitly displayed here.
Then, a power expansion in $\rho$, $I$ around $\uppi/2$ and $e$ of it is taken finally obtaining
\eqi
\ang{\Delta t^\mathrm{1pN}_D} \lb{bongo1} \simeq 6\,\Pb\,\ton{\rp{\mathcal{R}_g}{a}}\,\ton{\rp{R_\star}{a}}\,e\,\cos\omega\,\ton{-1 + 3\,e\,\sin \omega}.
\eqf
\rfr{bongo1}, which falls as $1/\sqrt{a}$, is valid up to terms of the order of $\mathcal{O}\ton{\rho},\, \mathcal{O}\ton{\delta^2},\,\mathcal{O}\ton{e^3}$,
where
\eqi
\delta \doteq I - \rp{\uppi}{2}
\eqf
is the departure--usually small for transiting exoplanets--of the orbital inclination $I$ from the ideal edge-on configuration.
In \rfr{bongo1},
\eqi
\mathcal{R}_g\doteq \rp{\mu}{c^2}
\eqf
is the system's characteristic pN gravitoelectric  length.
\textcolor{black}{It should be noted that, strictly speaking, \rfr{bongo1}, which vanishes for circular orbits, does not represent a genuine secular trend because of the trigonometric functions of $\omega$ entering it. Indeed, the pericentre changes secularly because of the pK effects considered in the present work; thus, in fact, \rfr{bongo1} is a long-period signal  modulated by the variation of $\omega$. However, from a practical point of view, \rfr{bongo1} can be considered as a secular signal because any realistically attainable observational time span will be always much shorter than the usually extremely long characteristic timescale of $\omega$. Also the other shifts worked out in Sections\,\ref{TLT}\,to\,\ref{TJ2} and in Sections\,\ref{ingr}\,to\,\ref{fwhm}  share the same feature.}
%
%
%
\subsubsection{The 1pN gravitomagnetic Lense-Thirring shift of $t_D$}\lb{TLT}
To the 1pN level, the gravitomagnetic LT net shift per orbit $\ang{\Delta t^\mathrm{LT}_D}$ of $t_D$  induced by the star's angular momentum ${\bds S}_\star$
is calculated as in Section\,\ref{TGE}. An approximated expression of it, valid up to terms of the order of $\mathcal{O}\ton{\rho},\, \mathcal{O}\ton{\delta},\,\mathcal{O}\ton{e^3}$, is
\eqi
\ang{\Delta t^\mathrm{LT}_D} \lb{pallina1} \simeq  16\,\uppi\,\mathcal{T}_\star^\mathrm{LT}\,\ton{\rp{R_\star}{a}}\,\ton{ {\bds{\hat{S}}}_\star\bds\cdot\bds{\hat{\mathrm{N}}} }\,{e\,\cos\omega\,\ton{1 - 3\,e\,\sin\omega}},
\eqf
where
\eqi
\mathcal{T}_\star^\mathrm{LT}\doteq \rp{S_\star}{c^2\,M_\star}
\eqf
is the characteristic gravitomagnetic timescale of a rotating body of mass $M_\star$ and angular momentum $S_\star$; for a Sun-like star it amounts to
\eqi
\mathcal{T}_\odot^\mathrm{LT} =1\times 10^{-6}\,\mathrm{s}.
\eqf
Furthermore,
$
\bds{\hat{\mathrm{N}}}
$
is the unit vector directed along the orbital angular momentum.
In transiting exoplanets, the stellar and orbital angular momenta are often approximately aligned, although spin-orbit misalignments are not infrequent \citep{2010ApJ...719..602S}, so that ${\bds{\hat{S}}}_\star$ and $\bds{\hat{\mathrm{N}}}$ can be assumed nearly parallel, and it can be posed
\eqi
{\bds{\hat{S}}}_\star\bds\cdot\bds{\hat{\mathrm{N}}}\simeq 1
\eqf
in \rfr{pallina1}.
%
\subsubsection{The shift of $t_D$ due to the quadrupole mass moment $J_2^\star$ of the star}\lb{TJ2}
The pK Newtonian net shift per orbit $\ang{\Delta t^{J_2^\star}_D}$ of $t_D$ induced by the stellar quadrupole mass moment $J_2^\star$, worked out as in Sections\,\ref{TGE} to \ref{TLT}, turns out to be too cumbersome to be explicitly displayed.
Then, a power expansion in $\rho,\,I,\,e$ is performed obtaining,  to the order of $\mathcal{O}\ton{\rho},\,\mathcal{O}\ton{\delta},\,\mathcal{O}\ton{e^3}$,
%
\eqi
\ang{\Delta t^{J_2^\star}_D}\lb{ttJ2}\simeq  \rp{3}{2}\,\Pb\,\ton{\rp{R_\star}{a}}^3\,J_2^\star\,\qua{-2 + 3\,\ton{{\bds{\hat{S}}}_\star\bds\cdot\bds{\hat{l}}}^2 + 3\,\ton{{\bds{\hat{S}}}_\star\bds\cdot\bds{\hat{m}}}^2}\,e\,\cos\omega\,\ton{1 - 3\,e\,\sin\omega}.
\eqf
In \rfr{ttJ2}, which falls as $1/a^{3/2}$, $\bds{\hat{m}}$
is the unit vector in the orbital plane perpendicular to the line of the nodes, and $\bds{\hat{l}}$
is the unit vector directed along the line of the nodes.
As already remarked, the star's spin is often nearly aligned with the orbital angular momentum, so that it can be posed
\eqi
{\bds{\hat{S}}}_\star\bds\cdot\bds{\hat{l}}\simeq {\bds{\hat{S}}}_\star\bds\cdot\bds{\hat{m}}\simeq 0
\eqf
in \rfr{ttJ2}.
\subsubsection{Numerical calculation}
\begin{figure}[H]
\centering
\centerline{
\vbox{
\begin{tabular}{c}
\epsfxsize= 8.5 cm\epsfbox{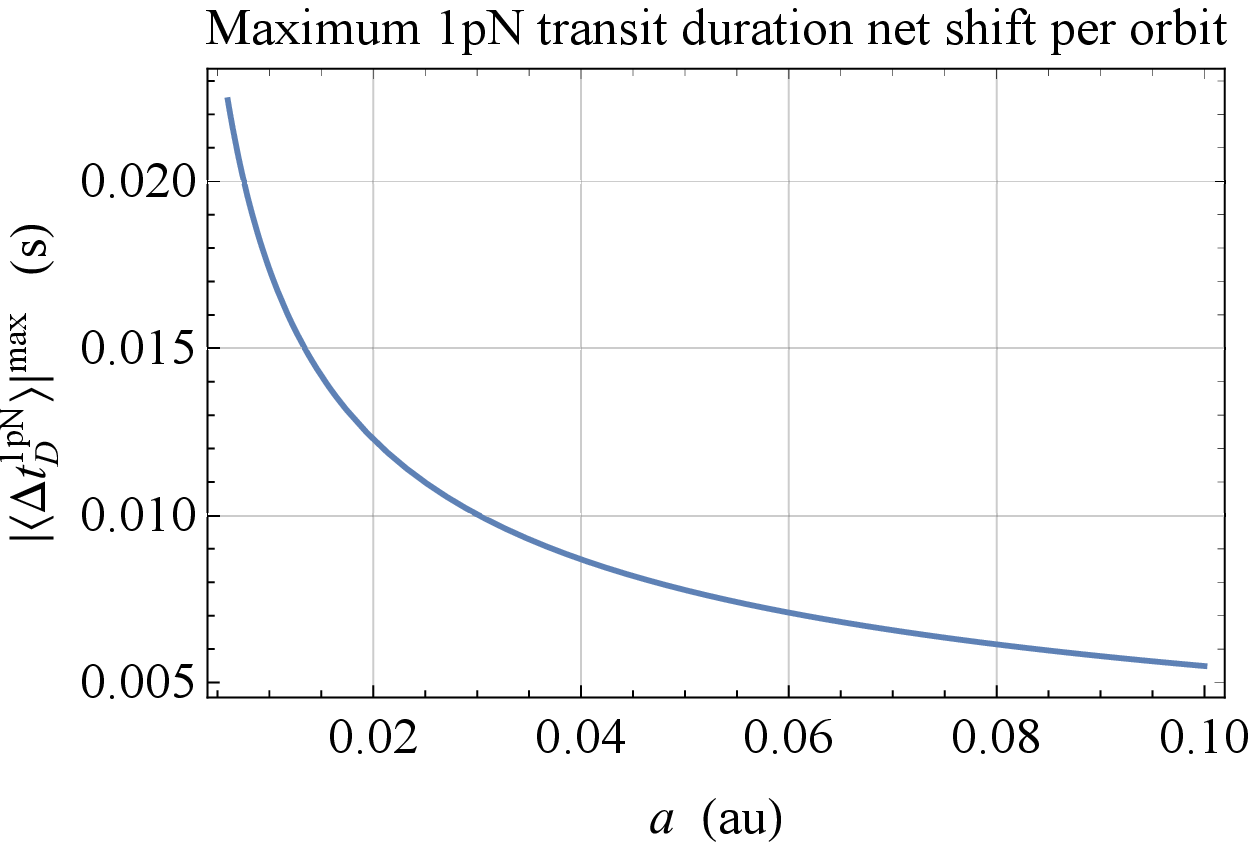}\\
\epsfxsize= 8.5 cm\epsfbox{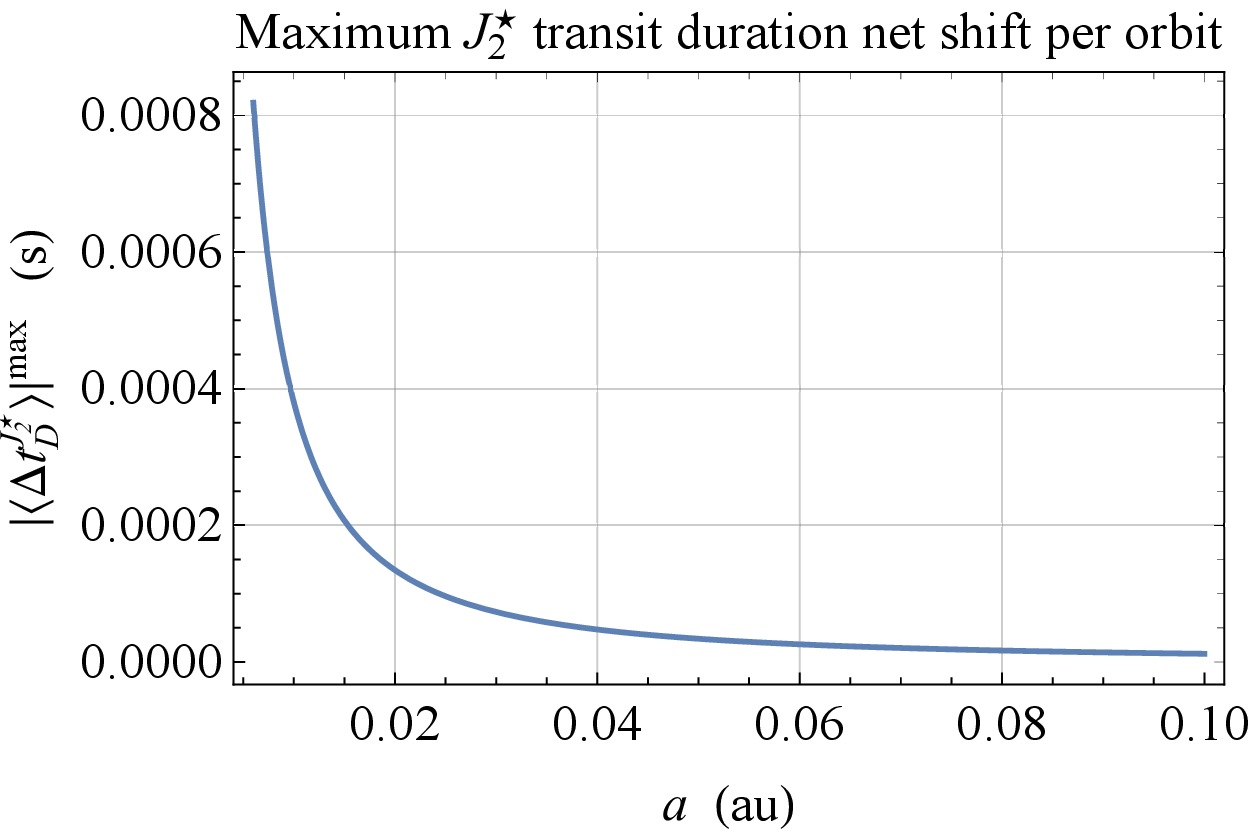}\\
\epsfxsize= 8.5 cm\epsfbox{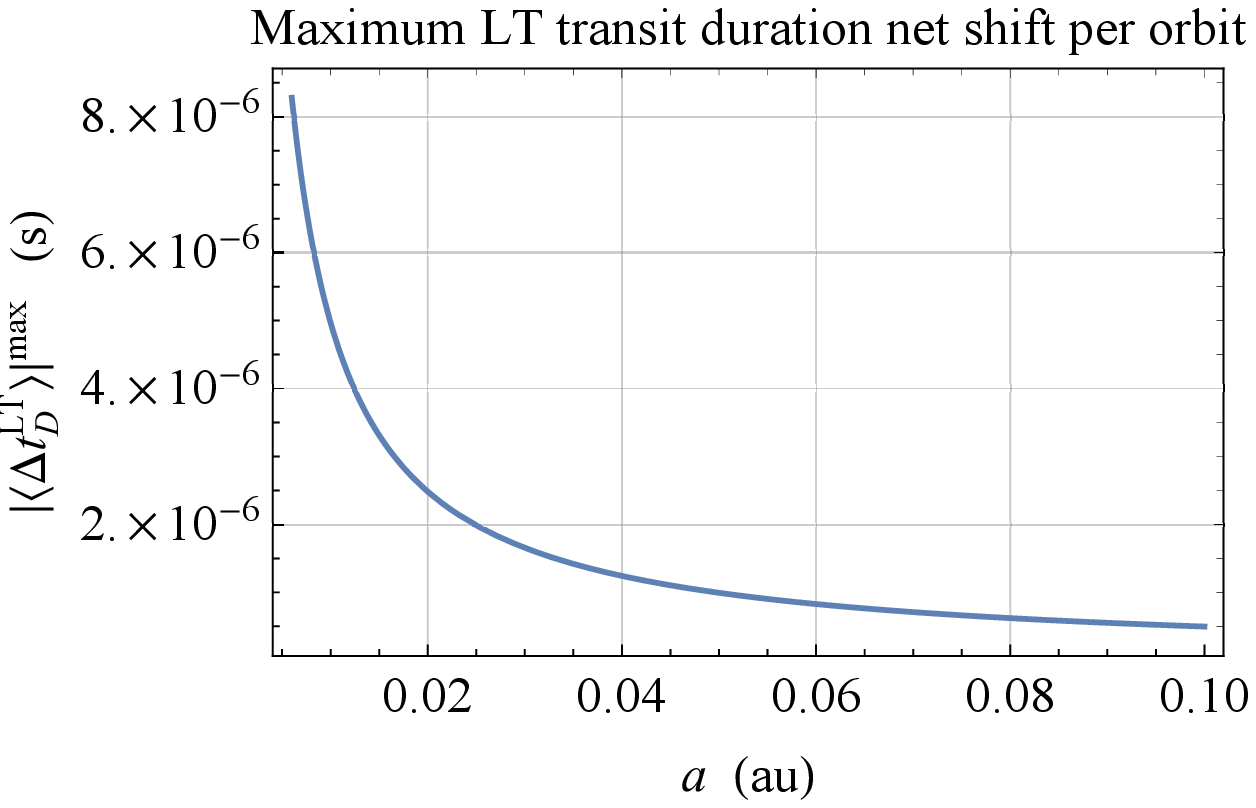}\\
\end{tabular}
}
}
\caption{
From the top to the bottom: maximum absolute values $\left|\ang{\Delta t_D}\right|^\mathrm{max}$ of the 1pN gravitoelectric, 1pN gravitomagnetic LT and classical quadrupole-induced net shifts per orbit of the total transit duration, in s,  as functions of the semimajor axis $a$, in au. They were calculated with \rfr{bongo1}, \rfr{pallina1} and \rfr{ttJ2} for a fictitious Sun-Jupiter system  by keeping the eccentricity fixed to $e=0.2$ and $\cos\omega = 1$.
}\label{fig1}
\end{figure}

\lb{misura}
Here, some numerical evaluations of the net shifts per orbit calculated in Sections\,\ref{TGE} to \ref{TJ2} are provided.
A Jupiter-like planet orbiting a Sun-type star along an elliptical orbit with $e=0.2$ and without spin-orbit misalignment is considered for different values of the semimajor axis $a$.
In Figure\,\ref{fig1}, the maxima of the absolute values $\left|\ang{\Delta t_D}\right|$ of \rfr{bongo1}, \rfr{pallina1} and \rfr{ttJ2}, in s, are plotted as functions of $a$.
It turns out that, for $a$ ranging from $0.006\,\mathrm{au}$ to $0.1\,\mathrm{au}$, the 1pN gravitoelectric net shift per orbit is of the order of \eqi
\ang{\Delta t_D^\mathrm{1pN}}\simeq 0.025-0.005\,\mathrm{s}.
\eqf The effect due to the stellar oblateness is two orders of magnitude smaller, ranging within
\eqi
\ang{\Delta t_D^{J_2^\star}}\simeq 0.0008-0.0001\,\mathrm{s}.
\eqf The smallest shift is the LT one; it is of the order of
\eqi
\ang{\Delta t_D^{LT}}\simeq \upmu \mathrm{s}.
\eqf

HD 286123 b \citep{2018MNRAS.477.2572B,2018AJ....156..127Y}, known also as K2--232b,  is a sub-Jovian transiting exoplanet which falls within the case of Figure\,\ref{fig1} since \citep{2018AJ....156..127Y}
$M_\star = 1.039\,M_\odot,\,R_\star=1.233\,R_\odot\,e = 0.255,\,a = 0.0991\,\mathrm{au},\,\omega \simeq 170^\circ$. Thus, its 1pN shift is of the order of
\eqi
\left|\ang{\Delta t_D^\mathrm{1pN}}\right|_\mathrm{HD\,286123\,b}\simeq 0.007\,\mathrm{s}.\lb{glo}
\eqf
The formal, statistical accuracy in measuring its total transit duration amounts to \citep{2018AJ....156..127Y}
\eqi
\sigma_{t_D}\simeq 0.0003\,\mathrm{d} = 28\,\mathrm{s}\lb{sigma}
\eqf
over a number of transits $N_\mathrm{tr}$ equal to
\eqi
N_\mathrm{tr}= 7
\eqf
since its orbital period is $\Pb = 11.16\,\mathrm{d}$, and it was monitored during 80 d \citep{2018AJ....156..127Y}.
It should be noted that \citet{2018AJ....156..127Y}, following the convention by \citet{2019arXiv190709480E}, use the symbol $T_{14}$ for the total transit duration.
This implies that, while the LT and quadrupole-driven shifts are undetectable in the foreseeable future, the 1pN gravitoelectric effect is, perhaps, not desperately so small not to fit into the measurability domain someday. Such a goal could be reached with extended data records covering much more transits and an improvement in $\sigma_{t_D}$ by $\simeq 1-2$ order of magnitude. For example, $N_\mathrm{tr} = 100$ transits would be observed over 3 yrs, yielding a total 1pN shift of $0.7\,\mathrm{s}$. In the meantime, it may likely not be unrealistic to expect some improvements  in $\sigma_{t_D}$; indeed, it generally improves with the number of transits $N_\mathrm{tr}$ as $\sqrt{N_\mathrm{tr}}$. Thus, for $N_\mathrm{tr} = 100$, $\sigma_{t_D}$ should reduce just by a factor of $10$. If one assumed a continuous monitoring over, say, 10 yrs, corresponding to $N_\mathrm{tr}\simeq 330$, the 1pN cumulative shift would amount to about $2.3\,\mathrm{s}$, while $\sigma_{t_D}$ would improve by a factor of $18$ corresponding to a measurement uncertainty as little as $1.6\,\mathrm{s}$; the measurability threshold would be, in principle, crossed. If it were possible to observe $N_\mathrm{tr} = 1000$ transits over about 30 yrs, the cumulative 1pN shift would become $7\,\mathrm{s}$, while the experimental accuracy would improve by a factor of 32 reaching the $0.8\,\mathrm{s}$ level; a $\simeq 11$ percent detection would be, at least in principle, possible. However, it must be remarked that the quoted errors are, in fact, just the formal, statistical ones; they do not account for several sources of systematics like, e.g., confusing time standards, neglecting star spots, neglecting clouds, etc.
\textcolor{black}{According to the online NASA Exoplanet Archive available at https://exoplanetarchive.ipac.caltech.edu, the best present-day accuracy in measuring the total transit duration of transiting exoplanets is of the order of
\eqi
\sigma_{t_D}\simeq 0.0004\,\mathrm{hr} = 1.4\,\mathrm{s}.\lb{sigma2}
\eqf
However, \rfr{sigma2} refers to planets moving along circular orbits, while \rfr{bongo1}, \rfr{pallina1} and \rfr{ttJ2} vanish for $e=0$.
}
\subsection{The ingress/egress transit duration $\uptau$}\lb{ingr}
Another characteristic timescale of transiting exoplanets which is measured in their data analyses is the ingress/egress transit duration \citep{2019arXiv190709480E}.

The total ingress duration can be defined as
\eqi
t_\mathrm{ing}\doteq t_\mathrm{II} - t_\mathrm{I}.
\eqf
With the same approximations used in Section\,\ref{trdr} for $t_D$, it can be calculated as
\eqi
t_\mathrm{ing} = \rp{\Delta\ell_\mathrm{I-II}}{v_\mathrm{mid}},\lb{tingo}
\eqf
where
\eqi
\Delta\ell_\mathrm{I-II}\doteq \ell_\mathrm{I} - \ell_\mathrm{II},\lb{Deltal}
\eqf
with
\begin{align}
\ell_\mathrm{I}& \doteq \ton{R_\star + R_\mathrm{p}}\,\cos\alpha_\mathrm{I}, \\ \nonumber \\
\ell_\mathrm{II}& \doteq \ton{R_\star - R_\mathrm{p}}\,\cos\alpha_\mathrm{II}.
\end{align}
The angles $\alpha_\mathrm{I},\,\alpha_\mathrm{II}$ can be obtained by imposing that
\begin{align}
\ton{R_\star + R_\mathrm{p}}\,\sin\alpha_\mathrm{I} & = y_\mathrm{p}^\mathrm{I}, \\ \nonumber \\
\ton{R_\star - R_\mathrm{p}}\,\sin\alpha_\mathrm{II} & = y_\mathrm{p}^\mathrm{II},
\end{align}
with the condition
\eqi
y_\mathrm{p}^\mathrm{I} = y_\mathrm{p}^\mathrm{II} = y_\mathrm{p}^\mathrm{mid} = \rp{a\,\ton{1-e^2}\,\cos I}{1 + e\,\sin\omega}.
\eqf
Thus, it is
\begin{align}
\sin\alpha_\mathrm{I} \lb{sina1}& = \rp{a\,\ton{1-e^2}\,\cos I}{R_\star\,\ton{1 + \rho}\,\ton{1 + e\,\sin\omega}}, \\ \nonumber \\
\sin\alpha_\mathrm{II} \lb{sina2}& = \rp{a\,\ton{1-e^2}\,\cos I}{R_\star\,\ton{1 - \rho}\,\ton{1 + e\,\sin\omega}},
\end{align}
where $\rho$ is given by \rfr{xi}.

By using \rfrs{tingo}{Deltal} along with \rfr{vmid} and \rfrs{sina1}{sina2}, one finally gets
\eqi
t_\mathrm{ing} = \rp{R_\star\,\sqrt{1-e^2}}{\nk\,a\,\sqrt{1 + 2\,e\,\sin\omega +  e^2}}\,\qua{\sqrt{\ton{1 + \rho}^2 - b^2} - \sqrt{\ton{1 - \rho}^2 - b^2}},\lb{Ting}
\eqf
where $b$ is defined as in \rfr{bi}.

Its relevant partial derivatives are
\begin{align}
\derp{t_\mathrm{ing}}{I} & = \rp{a\,\ton{1-e^2}^{5/2}\,\sin 2I}{2\,\nk\,R_\star\,\sqrt{1 + 2\,e\,\sin\omega + e^2}\,\ton{1 + e\,\sin\omega}^2}\,\qua{\rp{1}{\sqrt{\ton{1+\rho}^2 - b^2}}-\rp{1}{\sqrt{\ton{1-\rho}^2 - b^2}}}, \\ \nonumber \\
\derp{t_\mathrm{ing}}{\omega} \nonumber & = \rp{e\,R_\star\,\sqrt{1-e^2}\,\cos\omega}{\nk\,a\,\ton{1 + 2\,e\,\sin\omega + e^2}^{3/2}}\,\grf{
\sqrt{\ton{1-\rho}^2 - b^2} - \sqrt{\ton{1+\rho}^2 - b^2} + \right.\\ \nonumber \\
&\left. + \rp{a^2\,\ton{1-e^2}^2\,\cos^2 I\,\ton{1 + 2\,e\,\sin\omega + e^2}}{R^2_\star\,\ton{1 + e\,\sin\omega}^3}\qua{\rp{1}{\sqrt{\ton{1+\rho}^2 - b^2}}-\rp{1}{\sqrt{\ton{1-\rho}^2 - b^2}}}
}.
\end{align}

As far as the total egress duration  is concerned, it is defined as
\eqi
t_\mathrm{egr}\doteq t_\mathrm{IV} - t_\mathrm{III},
\eqf
and can be calculated as
\eqi
t_\mathrm{egr} = \rp{\Delta\ell_\mathrm{IV-III}}{v_\mathrm{mid}} \doteq \rp{\ell_\mathrm{IV}-\ell_\mathrm{III}}{v_\mathrm{mid}}.\lb{tegr}
\eqf
In \rfr{tegr}, it is
\begin{align}
\ell_\mathrm{III} & = \ton{R_\star - R_\mathrm{p}}\,\cos\alpha_\mathrm{III}, \\ \nonumber \\
\ell_\mathrm{IV} & = \ton{R_\star + R_\mathrm{p}}\,\cos\alpha_\mathrm{IV},
\end{align}
where the angles $\alpha_\mathrm{III},\,\alpha_\mathrm{IV}$ are defined from
\begin{align}
R_\star\,\ton{1 - \rho}\,\sin\ton{\uppi-\alpha_\mathrm{III}} & = y_\mathrm{p}^\mathrm{III}, \\ \nonumber \\
R_\star\,\ton{1 + \rho}\,\sin\ton{\uppi-\alpha_\mathrm{IV}} & = y_\mathrm{p}^\mathrm{IV}.
\end{align}
By imposing that
\eqi
y_\mathrm{p}^\mathrm{III} = y_\mathrm{p}^\mathrm{IV} = y_\mathrm{p}^\mathrm{mid} = \rp{a\,\ton{1-e^2}\,\cos I}{1 + e\,\sin\omega},
\eqf
one has
\begin{align}
\sin\alpha_\mathrm{III} \lb{sina3}& = \rp{a\,\ton{1-e^2}\,\cos I}{R_\star\,\ton{1 - \rho}\,\ton{1 + e\,\sin\omega}}, \\ \nonumber \\
\sin\alpha_\mathrm{IV} \lb{sina4}& = \rp{a\,\ton{1-e^2}\,\cos I}{R_\star\,\ton{1 + \rho}\,\ton{1 + e\,\sin\omega}}.
\end{align}
Thus, \rfr{tegr}, \rfr{vmid} and \rfrs{sina3}{sina4} yield for $t_\mathrm{egr}$ the same expression as \rfr{Ting} for $t_\mathrm{ing}$; in the following, they will be commonly denoted as $\uptau$, as in \citet{2019arXiv190709480E}.

By applying \rfr{metodo} to \rfr{Ting}, it is possible to calculate the net shift $\ang{\Delta \uptau}$ per orbit of the total ingress/egress duration for various pK accelerations.

Up to terms of the order of $\mathcal{O}\ton{\rho^3},\,\mathcal{O}\ton{\delta^2},\mathcal{O}\ton{e^3}$, the 1pN gravitoelectric effect turns out to be
\eqi
\ang{\Delta \uptau^\mathrm{1pN}}\simeq 6\,\Pb\,\ton{\rp{\mathcal{R}_g}{a}}\,\ton{\rp{R_\mathrm{p}}{a}}\,e\,\cos\omega\,\ton{-1 + 3\,e\,\sin\omega} = \rho\,\ang{\Delta t_D^\mathrm{1pN}}.\lb{tingGE}
\eqf

The 1pN gravitomagnetic LT net shift per orbit, up to terms of the order of
$\mathcal{O}\ton{\rho^3},\,\mathcal{O}\ton{\delta},\mathcal{O}\ton{e^3}$, is
\eqi
\ang{\Delta \uptau^\mathrm{LT}}\simeq 16\,\uppi\,\mathcal{T}_\star^\mathrm{LT}\,\ton{\rp{R_\mathrm{p}}{a}}\,\ton{{\bds{\hat{S}}}_\star\bds\cdot\bds{\hat{\mathrm{N}}}}\,e\cos\omega\,\ton{1 - 3\,e\,\sin\omega} = \rho\,\ang{\Delta t_D^\mathrm{LT}}.
\eqf

The classical pK effect due to the stellar quadrupole $J_2^\star$, up to terms of the order of
$\mathcal{O}\ton{\rho^3},\,\mathcal{O}\ton{\delta},\mathcal{O}\ton{e^3}$, reads
\begin{align}
\ang{\Delta \uptau^{J_2^\star}}\nonumber &\simeq \rp{3}{2}\,\Pb\,\ton{\rp{R_\mathrm{p}}{a}}\,\ton{\rp{R_\star}{a}}^2\,J_2^\star\,\qua{-2 + 3\,\ton{{\bds{\hat{S}}}_\star\bds\cdot\bds{\hat{l}}}^2 + 3\,\ton{{\bds{\hat{S}}}_\star\bds\cdot\bds{\hat{m}}}^2}\,e\cos\omega\,\ton{1 - 3\,e\,\sin\omega} = \\ \nonumber \\
&=\rho\,\ang{\Delta t_D^{J_2^\star}}.
\end{align}

The measurability of the 1pN gravitoelectric cumulative variation of the ingress/egress duration over the years is even more difficult than that for the  transit duration because, according to \rfr{tingGE},  it is reduced by a factor of $\simeq 0.1$ for a Sun-Jupiter system.
On the other hand, in the case of HD 286123 b,  the formal uncertainty in measuring the total ingress/egress duration is of the order of \citep{2018AJ....156..127Y}
\eqi
\sigma_{\uptau} = 0.00020-0.00047\,\mathrm{d}=17-40\,\mathrm{s}.
\eqf
\subsection{The full width at half maximum primary transit duration $t_H$}\lb{fwhm}
Another measurable characteristic timescale of the primary transit is the full width at half maximum primary transit duration \citep{2019arXiv190709480E}, which can be defined as
\eqi
t_H \doteq \ton{\rp{t_\mathrm{III} + t_\mathrm{IV}}{2}}-\ton{\rp{t_\mathrm{I} + t_\mathrm{II}}{2}};\lb{tH}
\eqf
\citet{2019arXiv190709480E} dub it as $T_{FWHM}$.
It can be calculated with the same approximations adopted in Section\,\ref{trdr} and Section\,\ref{ingr} as
\eqi
t_H = \rp{\ell_\mathrm{III} + \ell_\mathrm{IV} - \ell_\mathrm{I} -\ell_\mathrm{II}}{2\,\mathrm{v}_\mathrm{mid}}=\rp{R_\star\,\qua{\ton{1-\rho}\,\ton{\cos\alpha_\mathrm{III}-\cos\alpha_\mathrm{II}} +\ton{1+\rho}\,\ton{\cos\alpha_\mathrm{IV}-\cos\alpha_\mathrm{I}} }}{2\,\mathrm{v}_\mathrm{mid}}.\lb{tih}
\eqf
Since
\begin{align}
\cos\alpha_\mathrm{II} = -\cos\alpha_\mathrm{III}, \\ \nonumber \\
\cos\alpha_\mathrm{I} = -\cos\alpha_\mathrm{IV},
\end{align}
\rfr{tih}, by means of \rfr{vmid}, \rfr{sina3} and \rfr{sina4}, reduces to
\eqi
t_H = \rp{R_\star\,\sqrt{1-e^2}}{\nk\,a\,\sqrt{1 + 2\,e\,\sin\omega+e^2}}\,\qua{\sqrt{\ton{1-\rho}^2- b^2}+\sqrt{\ton{1+\rho}^2} - b^2},
\eqf
where $b$ is given by \rfr{bi}.
The relevant partial derivatives are
\begin{align}
\derp{t_H}{I} & =\rp{a\,\ton{1-e^2}^{5/2}\,\sin 2I}{2\,\nk\,R_\star\,\sqrt{1 + 2\,e\,\sin\omega + e^2}\,\ton{1 + e\,\sin\omega}^2}\,\qua{\rp{1}{\sqrt{\ton{1+\rho}^2 - b^2}}+\rp{1}{\sqrt{\ton{1-\rho}^2 - b^2}}}, \\ \nonumber \\
\derp{t_H}{\omega} \nonumber & = \rp{e\,R_\star\,\sqrt{1-e^2}\,\cos\omega}{\nk\,a\,\ton{1 + 2\,e\,\sin\omega + e^2}^{3/2}}\,\grf{
-\sqrt{\ton{1-\rho}^2 - b^2} - \sqrt{\ton{1+\rho}^2 - b^2} + \right.\\ \nonumber \\
&\left. + \rp{a^2\,\ton{1-e^2}^2\,\cos^2 I\,\ton{1 + 2\,e\,\sin\omega + e^2}}{R^2_\star\,\ton{1 + e\,\sin\omega}^3}\qua{\rp{1}{\sqrt{\ton{1+\rho}^2 - b^2}}+\rp{1}{\sqrt{\ton{1-\rho}^2 - b^2}}}
}.
\end{align}

The 1pN gravitoelectric net shift per orbit, up to terms of the order of $\mathcal{O}\ton{\rho^2},\,\mathcal{O}\ton{\delta^2},\,\mathcal{O}\ton{e^3}$, is
\eqi
\ang{\Delta t_H^\mathrm{1pN}} = \ang{\Delta t_D^\mathrm{1pN}}.\lb{nuu}
\eqf

The 1pN gravitomagnetic LT and the quadrupole-driven net shifts per orbit, up to terms of the order of $\mathcal{O}\ton{\rho^2},\,\mathcal{O}\ton{\delta},\,\mathcal{O}\ton{e^3}$, turns out to be
\begin{align}
\ang{\Delta t_H^\mathrm{LT}} &= \ang{\Delta t_D^\mathrm{LT}},\\ \nonumber \\
\ang{\Delta t_H^{J_2^\star}} &= \ang{\Delta t_D^{J_2^\star}}.
\end{align}

About the measurability of \rfr{nuu}, \citet{2018AJ....156..127Y} report for  HD 286123 b the formal error
\eqi
\sigma_{t_H} \simeq 0.0002\,\mathrm{d} = 17\,\mathrm{s}.
\eqf
Thus, the situation seems slightly more favorable than that outlined in Section\,\ref{misura} for $t_D$.
\subsection{The time of inferior conjunction $t_\mathrm{cj}$}\lb{TC}
Another measurable quantity in transiting exoplanets is the time of inferior conjunction $t_\mathrm{cj}$ \citep{2019arXiv190709480E}. Its explicit expression can be obtained as follows.

The standard textbook expression of time $t$ as a function of the true anomaly $f$ reads \citep{2005som..book.....C}
\eqi
t = t_p + \rp{1}{\nk}\,\grf{2\,\arctan\qua{\sqrt{\rp{1-e}{1+e}}\,\tan\ton{\rp{f}{2}}}-\rp{e\,\sqrt{1-e^2}\,\sin f}{1 + e\,\cos f}},\lb{ti}
\eqf
where $t_p$ is the time of passage at pericentre; \rfr{ti} can be straightforwardly derived by integrating \rfr{dfdt} with respect to the true anomaly from 0, which is taken at pericentre, to an arbitrary value $f$.

In order to obtain $t_\mathrm{cj}$, \rfr{ti} has to be calculated  with the value of the true anomaly at midtransit given by \rfr{fmid}. Thus, one has
\eqi
t_\mathrm{cj} = t_p + \rp{1}{\nk}\,\grf{2\,\arctan\qua{\sqrt{\rp{1-e}{1+e}}\,\tan\ton{\rp{\uppi}{4}-\rp{\omega}{2}}}-\rp{e\,\sqrt{1-e^2}\,\cos \omega}{1 + e\,\sin\omega}}.\lb{tconj}
\eqf

\Rfr{tconj} can be used in \rfr{DF} to work out the net shifts per orbit $\ang{\Delta t_\mathrm{cj}}$ induced by the various pK accelerations;
the relevant partial derivative turns out to be
\eqi
\derp{t_\mathrm{cj}}{\omega} = -\rp{\ton{1-e^2}^{3/2}}{\nk\,\ton{1 + e\,\sin\omega}^2}.
\eqf
In order to calculate the full variation of $t_\mathrm{cj}$, also the change of $t_p$ has to be calculated as
\eqi
\Delta t_p = -\rp{\Delta\eta}{\nk},
\eqf
where $\Delta\eta$ is worked out by means of \rfr{Deta} and \rfr{detadt}.

Interestingly, it turns out that the 1pN gravitoelectric shift has a non-vanishing term to the order zero in the eccentricity which increases with the distance from the star as $\sqrt{a}$; it is
\begin{align}
\ang{\Delta t_\mathrm{cj}^\mathrm{1pN}} \nonumber \lb{conjGE}&= 3\,\Pb\,\ton{\rp{\mathcal{R}_g}{a}}\,\rp{4 + e^2 - 3\,\nu -
e\,\ton{-5 + 3\,\nu}\,\sin\omega\,\ton{2 + e\,\sin\omega}}{\sqrt{1-e^2}\,\ton{1 + e\,\sin\omega}^2} \simeq \\ \nonumber \\
&\simeq 3\,\Pb\,\ton{\rp{\mathcal{R}_g}{a}}\,\ton{4 - 3\,\nu} + \mathcal{O}\ton{e},
\end{align}
where
\eqi
\nu\doteq \rp{M_\mathrm{A}\,M_\mathrm{B}}{M_\mathrm{tot}^2}.
\eqf
The term of zero order in $e$ of \rfr{conjGE} is  a secular one since only $a$, which does not experience net variations per orbit, enters it.

The 1pN gravitomagnetic LT net shift per orbit is
\eqi
\ang{\Delta t_\mathrm{cj}^\mathrm{LT}} = \rp{4\,\uppi\,\mathcal{T}_\star^\mathrm{LT}\,\qua{2\,\ton{\bds{\hat{S}}_\star\bds\cdot\bds{\hat{\mathrm{N}}}} + \ton{\bds{\hat{S}}_\star\bds\cdot\bds{\hat{m}}}\,\cot I}}{\ton{1 + e\,\sin\omega}^2}\simeq 8\,\uppi\,\mathcal{T}_\star^\mathrm{LT}.\lb{conjLT}
\eqf
The last term in \rfr{conjLT}, which is independent of $a$, holds for spin--orbit alignment and for circular orbits.

The stellar quadrupole mass moment $J_2^\star$ induces a net shift per orbit given by
\begin{align}
\ang{\Delta t_\mathrm{cj}^{J_2^\star}} \nonumber \lb{conjJ2}&= -\rp{3}{8}\,\Pb\,\ton{\rp{R_\star}{a}}^2\,J_2^\star\,\rp{1}{\ton{1-e^2}^{3/2}\,\ton{1 + e\,\sin\omega}^2}\,\grf{
4\,\ton{1 - e^2}\,\ton{{\bds{\hat{S}}}_\star\bds\cdot\bds{\hat{\mathrm{N}}}}\,\ton{{\bds{\hat{S}}}_\star\bds\cdot\bds{\hat{m}}}\,\cot I  + \right.\\ \nonumber \\
\nonumber &\left. + \qua{-2 + 3\,\ton{{\bds{\hat{S}}}_\star\bds\cdot\bds{\hat{l}}}^2 + 3\,\ton{{\bds{\hat{S}}}_\star\bds\cdot\bds{\hat{m}}}^2}\,\ton{-4 + e^2 + e^2\,\cos 2\omega - 4\,e\,\sin\omega}} \simeq \\ \nonumber \\
&\simeq \rp{3}{2}\,\Pb\,\ton{\rp{R_\star}{a}}^2\,J_2^\star\,\qua{-2 + 3\,\ton{{\bds{\hat{S}}}_\star\bds\cdot\bds{\hat{l}}}^2 + 3\,\ton{{\bds{\hat{S}}}_\star\bds\cdot\bds{\hat{m}}}^2} +\mathcal{O}\ton{e}.
\end{align}
\Rfr{conjJ2}, which falls as $1/\sqrt{a}$, is valid for edge-on orbital configurations; for no spin--orbit misalignment, its last term reduces to
\eqi
\ang{\Delta t_\mathrm{cj}^{J_2^\star}}\simeq -3\,\Pb\,\ton{\rp{R_\star}{a}}^2\,J_2^\star + \mathcal{O}\ton{e}.
\eqf

\begin{figure}[H]
\centering
\centerline{
\vbox{
\begin{tabular}{c}
\epsfxsize= 8.5 cm\epsfbox{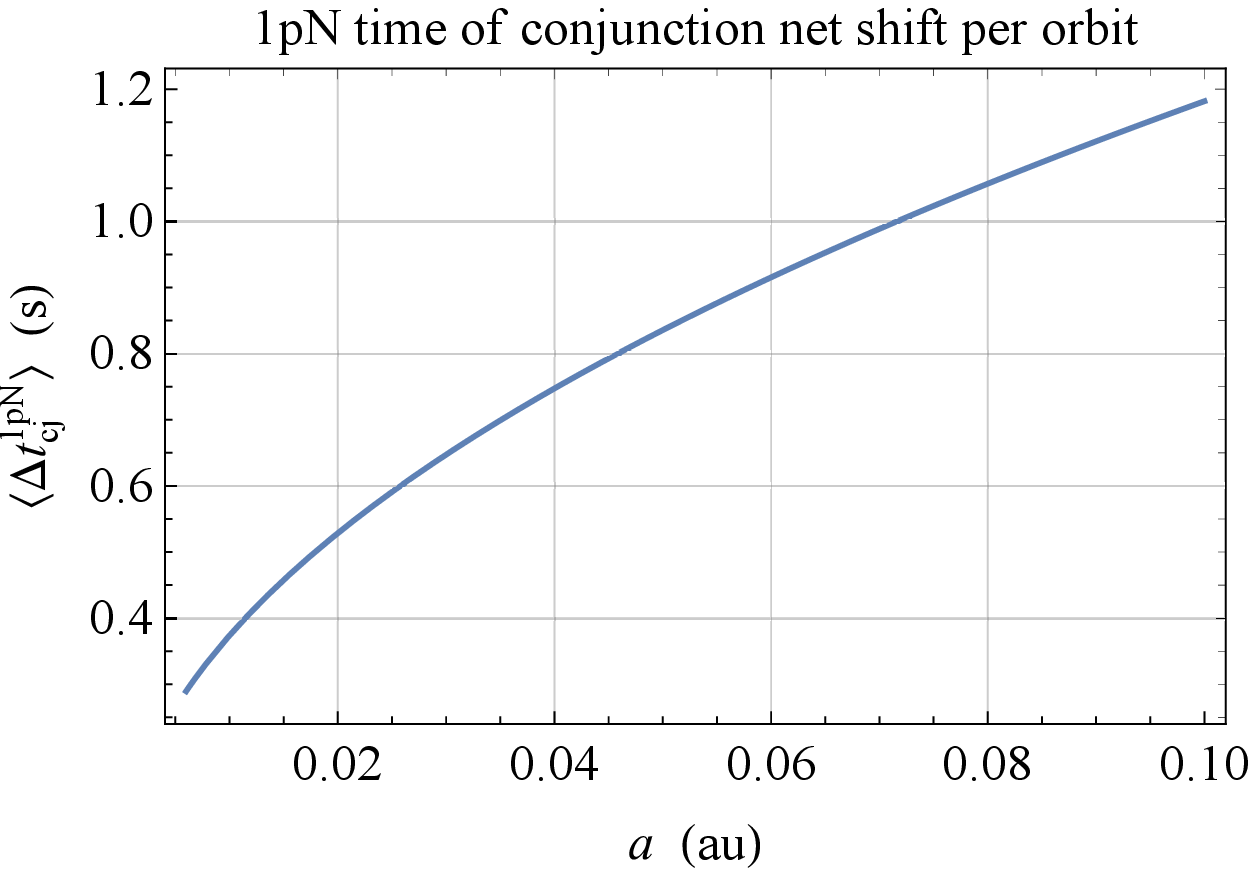}\\
\epsfxsize= 8.5 cm\epsfbox{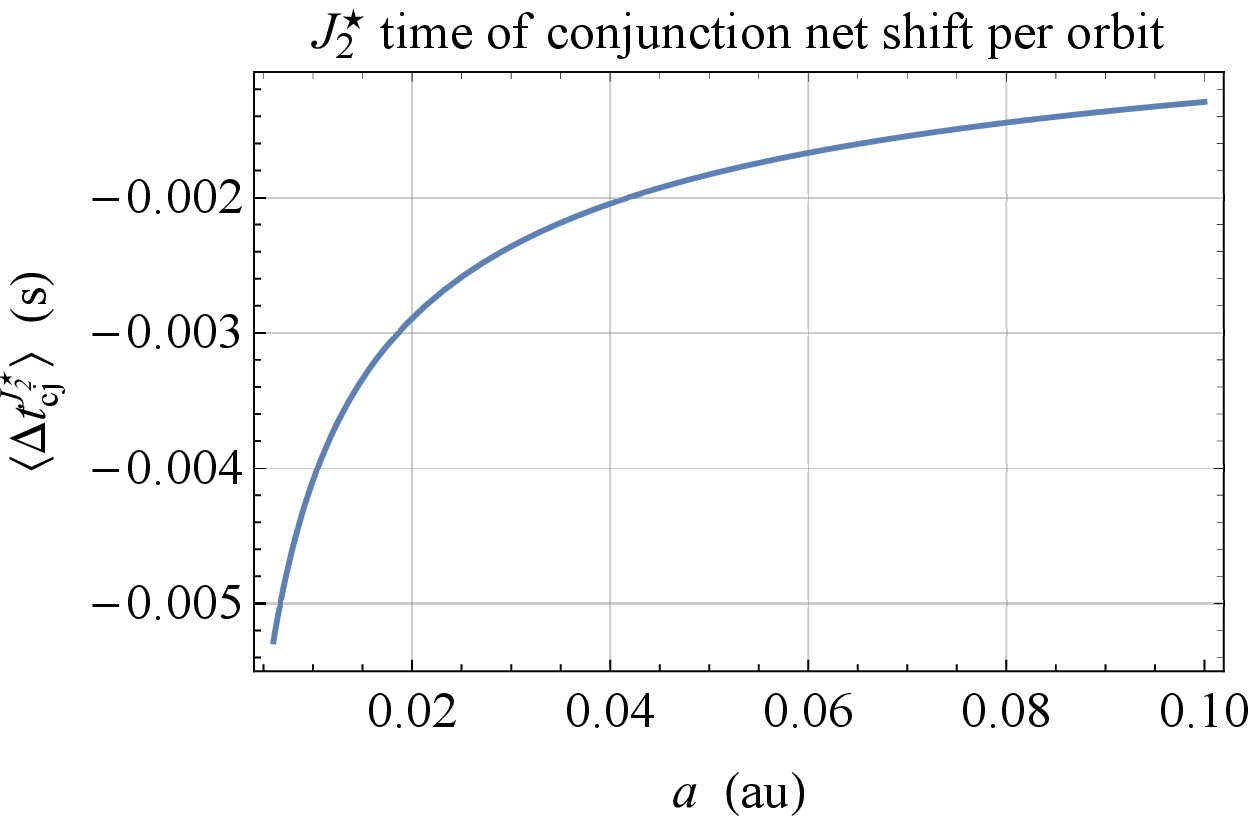}\\
\epsfxsize= 8.5 cm\epsfbox{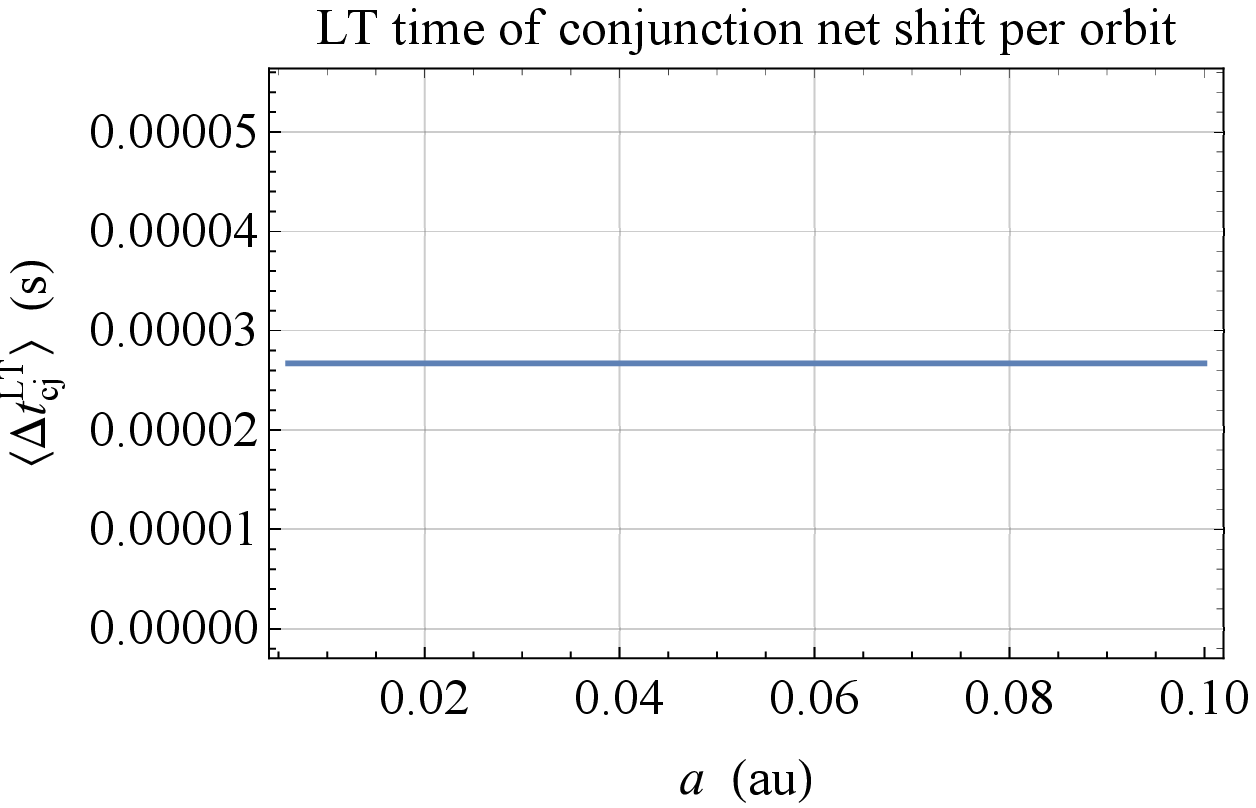}\\
\end{tabular}
}
}
\caption{
From the top to the bottom: 1pN gravitoelectric, 1pN gravitomagnetic LT and classical quadrupole-induced net shifts per orbit of the time of inferior conjunction $t_\mathrm{cj}$, in s,  as functions of the semimajor axis $a$, in au. They were calculated with \rfr{conjGE}, \rfr{conjLT} and \rfr{conjJ2} for a fictitious Sun-Jupiter system  by keeping the eccentricity fixed to $e=0.2$.
}\label{fig2}
\end{figure}

In Figure\,\ref{fig2}, the plots of \rfrs{conjGE}{conjJ2} for the fictitious Sun-Jupiter system of Figure\,\ref{fig1} are displayed.
For $a$ ranging from $0.006$ to $0.1\,\mathrm{au}$, the 1pN shift raises from $0.3$ to $1.2\,\mathrm{s}$, while the magnitude of the quadrupole-driven effect passes from 5 to $1\,\mathrm{ms}$; the 1pN LT shift amounts to about $30\,\upmu\mathrm{s}$.

The 1pN gravitoelectric net shift per orbit of HD 286123 b amounts to
\eqi
\ang{\Delta t_\mathrm{cj}^\mathrm{1pN}} = 1.3\,\mathrm{s}.
\eqf
The reported formal uncertainty on the time of inferior conjunction is \citep{2018AJ....156..127Y}
\eqi
\sigma_{t_\mathrm{cj}} \simeq 0.00004\,\mathrm{d} = 3.6\,\mathrm{s}
\eqf
after $N_\mathrm{tr}=7$ transits which correspond to a cumulative 1pN shift of $9.1\,\mathrm{s}$. Thus, it should already be within the detectability threshold, at least in principle. The exquisite--although merely statistical--accuracy with which $t_\mathrm{cj}$ can be measured for this planet opens interesting perspectives about the possibility of measuring the largest pN effect. Indeed, after 3 yrs and $N_\mathrm{tr}=100$ transits, the cumulative pN shift  becomes as large as $130\,\mathrm{s}$, while the measurement error should be reduced down to $\simeq 0.3\,\mathrm{s}$. If it were possible to observe $N_\mathrm{tr} = 330$ transits over 10 yrs, the error should become as little as $\simeq 0.2\,\mathrm{s}$, while the total pN variation would be $429\,\mathrm{s}$: such figures correspond to a $\simeq 5\times 10^{-4}$ relative accuracy in a possible measurement of it. With $N_\mathrm{tr}=1000$ transits over 30 yrs, it may reach, at least in principle, the $\simeq 8\times 10^{-5}$ level. About the realistic obtainable accuracy, the same caveats pointed out at the end of Section\,\ref{misura} hold also here.
\section{The secondary eclipse: when the planet is occulted by the star}\lb{secondary}
When the planet passes behind the star, it is said that the secondary transit, or the secondary eclipse, occurs \citep{Winn2011}.
The calculation of the measurable timescales characterizing the planet's occultation, i.e. the total eclipse duration, the ingress/egress eclipse duration the full width at half maximum primary transit duration and the time of superior conjunction \citep{2019arXiv190709480E},  goes as for their primary transit's counterparts in Sections\,\ref{trdr}\,to\,\ref{TC},
with the difference that the true anomaly at midtransit $f_\mathrm{mid}$ is, now, shifted by $\uppi$ with respect to  \rfr{fmid}. Thus,
\eqi
\cos f_\mathrm{mid} = \cos\ton{\rp{3}{2}\,\uppi-\omega} = -\sin\omega
\eqf
yielding
\eqi
\mathrm{v}_\mathrm{mid} = \rp{\nk\,a}{\sqrt{1-e^2}}\,\sqrt{1 -2\,e\,\sin\omega + e^2},\lb{vmidocc}
\eqf
and
\eqi
b \doteq \rp{a\,\ton{1-e^2}\,\cos I}{R_\star\,\ton{1 - e\,\sin\omega}}.\lb{bocc}
\eqf

As far as $\ang{\Delta t_D}$, $\ang{\Delta\uptau}$ and $\ang{\Delta t_H}$ are concerned, it turns out that, in the case of the secondary transit, their expressions differ from those for the primary transit just for a minus sign in front of the terms of the order of $\mathcal{O}\ton{e}$. Instead, the 1pN gravitoelectric net shift of the time of superior conjunction is, to the zero order in $e$, equal to that of inferior conjunction being the terms of the order of $\mathcal{O}\ton{e^{2k+1}}$ different by a minus sign. The 1pN gravitomagnetic LT shift $\ang{\Delta t^\mathrm{LT}_\mathrm{cj}}$ and the quadrupole-induced variation $\ang{\Delta t^{J_2^\star}_\mathrm{cj}}$ turn out to be identical for both the conjunctions.

In view of a comparison with observations in the case of the secondary eclipse of HD 286123 b, \citet{2018AJ....156..127Y}, following the conventions by \citet{2019arXiv190709480E}, use the symbols  $T_{S,14}$, $\uptau_S$, $T_{S,FWHM}$ and $T_S$ instead of $t_D$, $\uptau$, $t_H$ and $t_\mathrm{cj}$, respectively. The formal uncertainties reported by \citet{2018AJ....156..127Y} are worse than those for the primary eclipse, making, thus, even more difficult the detection of the 1pN effects in the case of the secondary transit. Indeed, they are \citep{2018AJ....156..127Y}
\begin{align}
\sigma_{t_D} & \simeq 0.01\,\mathrm{d} = 864\,\mathrm{s}, \\ \nonumber \\
\sigma_\uptau & \simeq 0.001\,\mathrm{d} = 86\,\mathrm{s}, \\ \nonumber \\
\sigma_{t_H} &\simeq 0.01\,\mathrm{d} = 864\,\mathrm{s}, \\ \nonumber \\
\sigma_{t_\mathrm{cj}} &\simeq 0.2\,\mathrm{d} = 4.8\,\mathrm{hrs} = 17280\,\mathrm{s}.
\end{align}
\section{Summary and conclusions}\lb{concludi}
A general analytical approach to perturbatively calculate the net shifts per orbit $\ang{\Delta F}$ induced by an extra-acceleration $\bds A$ of whatsoever physical origin on any observable $\mathfrak{O}$ which can be modeled as an explicit function $F$ of the Keplerian orbital elements $a,\,e,\,I,\,\Omega,\,\omega,\,f$  was devised.

It was applied to the characteristic timescales $\grf{t_\mathrm{trn}}$ which are usually measured in the data reductions of the light curves of transiting exoplanets, i.e.  the total transit duration $t_D$, the ingress/egress transit duration $\uptau$, the full width at half maximum primary transit duration $t_H$, and the time(s) of conjunction(s) $t_\mathrm{cj}$ in order to calculate their averaged variations caused by the pK accelerations due to the star's quadrupole mass moment $J_2^\star$ and to the pN gravitoelectric (Schwarzschild) and gravitomagnetic (Lense-Thirring) components of the stellar gravitational field. The calculation was performed for both the primary and the secondary transits.

The resulting formulas for $t_D,\,\uptau,\,t_H$, valid for small departures from the ideal edge-on orbital configuration  and to the second order in the eccentricity $e$, were calculated for a fictitious Sun-Jupiter system and plotted as functions of the semimajor axis $a$ for a given value of the eccentricity ($e=0.2$). It turns out that, for $a$ ranging from $0.006$ to $0.1\,\mathrm{au}$, the largest effects are the 1pN gravitoelectric ones whose maximum amplitudes per orbit for $t_D$ and $t_H$  are of the order of $0.025-0.005\,\mathrm{s}$. Their measurability depends on the number of transits $N_\mathrm{tr}$ available and on the experimental accuracy $\sigma$  which improves roughly by a factor of $\sqrt{N_\mathrm{tr}}$. In the case of HD 286123 b, a sub-Jovian planet orbiting its Sun-like star in $11.6\,\mathrm{d}$ along an elliptical orbit with $e = 0.2555$, the current error in measuring the total transit duration is $\sigma_{t_D} = 28\,\mathrm{s}$ over $N_\mathrm{tr} = 7$ transits. If $N_\mathrm{tr} = 1000$ transits were available, corresponding to a continuous monitoring of about $30\,\mathrm{yrs}$, a $\simeq 11$ percent detection of the cumulative 1pN variation of $t_D$ might be, in principle, feasible. The situation for the time of inferior conjunction $t_\mathrm{cj}$ is much more promising. Also their pK net variations per orbit were plotted for the same fictitious Sun-Jupiter system by obtaining that the 1pN gravitoelectric effect can reach the $\simeq 1\,\mathrm{s}$ level for $a=0.1\,\mathrm{au}$, while the $J_2^\star$-induced shift is of the order of  $\simeq \mathrm{ms}$; the 1pN gravitomagnetic shift is as little as $\simeq 10\,\upmu\mathrm{s}$. About HD 286123 b, since $\sigma_{t_\mathrm{cj}} = 3.6\,\mathrm{s}$ over $N_\mathrm{tr} = 7$ transits, after 30 yrs and  $N_\mathrm{tr} = 1000$ transits, a  measurement of the cumulative 1pN variation of $t_\mathrm{cj}$ at a  $\simeq 8\times 10^{-5}$  level of relative accuracy might be possible, at least in principle. To this aim, it is important to stress that the quoted measurement errors are just the formal, statistical ones; systematics like, e.g., confusing time standards, neglecting star spots, neglecting clouds, etc. would likely deteriorate the finally attainable accuracy.

In principle, the strategy outlined here can be applied also to modified models of gravity in order to calculate their impact on the relevant timescales of transiting exoplanets and preliminarily constraining their key parameters from a comparison of the resulting net shifts with the current measurement errors. It will be the subject of a forthcoming paper.
\section*{Acknowledgements}
I am grateful to Jason Eastman, Joseph Rodriguez, Gregory Gilbert and Eric B. Ford for useful information and stimulating discussions.
\section*{Data availability}
No new data were generated or analysed in support of this research.
\bibliography{exopbib,PXbib,IorioFupeng}{}


\end{document}